\documentclass[12pt]{article}
\usepackage{a4wide}
\usepackage{amssymb}
\begin{document}
{\renewcommand{\thefootnote}{\fnsymbol{footnote}}
\hfill  IGC--11/12--1\\
\medskip
\begin{center}
{\LARGE  Generalized uncertainty principles\\ and localization in discrete space}\\
\vspace{1.5em}
Martin Bojowald$^1$\footnote{e-mail address: {\tt bojowald@gravity.psu.edu}}
and Achim Kempf$^2$\footnote{e-mail address: {\tt akempf@uwaterloo.ca}}
\\
\vspace{0.5em}
$^1$Institute for Gravitation and the Cosmos,
The Pennsylvania State
University,\\
104 Davey Lab, University Park, PA 16802, USA\\
\vspace{0.5em}
$^2$Dept. of Applied Mathematics, University of Waterloo \\
Waterloo, Ontario, N2L 3G1, Canada 
\vspace{1.5em}
\end{center}
}

\setcounter{footnote}{0}

\begin{abstract}
  Generalized uncertainty principles are able to serve as useful
  descriptions of some of the phenomenology of quantum gravity
  effects, providing an intuitive grasp on non-trivial space-time
  structures such as a fundamental discreteness of space, a universal
  band\-limit or an irreducible extendedness of elementary
  particles. In this article, uncertainty relations are derived by a
  moment expansion of states for quantum systems with a discrete
  coordinate, and correspondingly a periodic momentum. Corrections to
  standard uncertainty relations are found, with some similarities but
  also key differences to what is often assumed in this context. The
  relations provided can be applied to discrete models of matter or
  space-time, including loop quantum cosmology.
\end{abstract}

\section{Introduction}

If space is discrete, the form of its underlying structure should influence the
general properties of position and momentum measurements and therefore their 
fundamental uncertainty relations. Compared with
standard quantum mechanics, there may be additional limitations to the
precision of measurements, as they can often be captured in
generalized uncertainty principles
\cite{ShortDistance,ShortDistanceThree,NonLocalGUP}. Such
modifications are bound to arise because the momentum, on a discrete space,
is no longer defined in all situations; in general, it must be
replaced by finite translation operators for displacements of at least
the lattice spacing. On scales larger than the lattice spacing, one
may introduce an approximate momentum operator, just as one can define
approximate plane waves of wavelength larger than the
spacing. However, as the wave length approaches the discreteness
scale, the underlying structure becomes noticeable and deviations from
standard properties of momentum arise. 

In the context of the low-energy regime of various approaches to
quantum gravity it is therefore of interest to explore the
consequences of spatial discreteness for the basic uncertainty
relations.  In this paper, we present a systematic method to compute
the leading corrections to the position and momentum uncertainty
relations for discrete spaces. Differences to some common assumptions
about such principles are pointed out. We begin this article with a
brief review of the mathematical structures involved in discrete
matter systems on the one hand, and some approaches to quantum gravity
on the other. Our discussion will focus on localization, in the sense
of minimizing fluctuations in position, and we will study uncertainty
principles without needing to refer to specific representations.  In
the main part of this article, Section~\ref{s:Circle}, we will then
systematically derive the generalized uncertainty principle for a
discrete system.

\section{Spatial discreteness}
\label{s:Basis}

There are numerous examples of discrete structures in physical models, such as
crystals that have periodic potentials. As an illustration, let us
consider the 1-dimensional quantum mechanical system of Bloch states. For
wavelike excitations of a length well above the periodicity of the crystal,
one may start with free scattering states $\exp(ikq)$ in the position
representation, whose energy is $E(k)=\hbar^2k^2/2m$ if they represent
particles of mass $m$. These states are no longer energy eigenstates if the
particles move in a non-trivial periodic potential $V(q)$ with
$V(q+q_0)=V(q)$, where $q_0$ is the periodicity. We decompose the set of plane
waves into sectors labeled by a real number $\epsilon\in [0,2\pi)$ in
one-to-one correspondence with wave functions on the finite interval $[0,q_0]$
subject to the ``almost periodic'' boundary condition
$\psi(q+q_0)=e^{i\epsilon} \psi(q)$. Square integrable functions satisfying
these boundary conditions define the Hilbert spaces ${\cal
  H}_{\epsilon}$. Parameterized by $\epsilon$ for all the sectors, momentum
eigenstates are then
\begin{equation}
 \psi_n^{(\epsilon)}(q)= \exp(i\mu_n^{(\epsilon)}q)
\end{equation}
where for all integers $n$,
\begin{equation} \label{mu}
 \mu_n^{(\epsilon)}:= \frac{2\pi n+\epsilon}{q_0}
\end{equation}
is proportional to the momentum eigenvalues
\begin{equation}
 p_n^{(\epsilon)}= \hbar \mu_n^{(\epsilon)}\,.
\end{equation}

For each fixed $\epsilon$, these are the discrete momentum eigenvalues of a
particle on a circle with $e^{i\epsilon}$-periodicity, and together, for all
$\epsilon$, they fill the whole real line.  In this heuristic way, the
continuous momentum spectrum for a particle in the periodic potential is
recovered. This statement is heuristic because the Hilbert spaces ${\cal
  H}_{\epsilon}$ are all different as function spaces and independent for
different $\epsilon$, and a wave function $\psi_n^{(\epsilon)}(q)$ would not
be normalizable in the usual continuum Hilbert space $L^2({\mathbb R},{\rm
  d}q)$. One may view the Hilbert spaces of different $\epsilon$ as
superselection sectors in the direct sum $\bigoplus_{\epsilon}{\cal
  H}_{\epsilon}$: One would consider all states as lying in the same Hilbert
space, but allow superpositions only of states within the same ${\cal
  H}_{\epsilon}$. (The full direct-sum Hilbert space is non-separable.)

In contrast to the momentum spectrum, the energy spectrum in a given
periodic potential $V(q)$, while continuous, need not fill the whole
real line. By solving the energy eigenvalue equation for each
$\epsilon$, $\hat{H}\psi_k^{(\epsilon)}=
E^{(\epsilon)}(k)\psi_k^{(\epsilon)}$ where $\psi_k^{(\epsilon)}$ is
subject to the almost-periodicity condition, one obtains a function
$E^{(\epsilon)}(k)$. Combining all values for the different $\epsilon$
in general leaves out some real numbers which are not realized as an
energy eigenvalue in the periodic potential, and the band structure of
excitation spectra emerges. 

Functional analytically, the differential operator $-\hbar^2\partial_q^2/2m +
V(q)$, when considered on the finite interval of one periodicity length, 
  becomes self-adjoint once suitable boundary conditions are imposed. Its
spectrum depends on the boundary conditions. The operator has deficiency
indices (2,2) and thus possesses a family of self-adjoint extensions
parameterized by U(2). Our previous boundary conditions $\psi(q+q_0) = \psi(q)
e^{i\epsilon}$ combined with $\psi'(q+q_0) = \psi'(q) e^{i\epsilon}$ amount to
a subgroup ${\rm U}(1)\subset {\rm U}(2)$. For each choice of such a boundary
condition, i.e, for each choice of $\epsilon \in [0, 2\pi)$, we obtain a
different self-adjoint extension $\hat{H}_\epsilon$, each possessing its own
spectrum and eigenvectors.  Each set of eigenvectors spans the same Hilbert
space of square integrable functions over the interval, and the union of these
spectra forms the bands.

Clearly, the underlying periodicity of the crystal, by leading to the band
structure, has direct implications for the dynamics, which allows one to probe
underlying properties of $V(q)$ in experiments. In low-energy experiments,
distance scales larger than the spatial periodicity can easily be probed and
described perturbatively, for instance by corrected dispersion relations
taking into account the microstructure. Of interest in the present context is
the fact that a discrete structure arises in momentum space as a consequence
of periodicity in position space.

Some approaches to quantum cosmology, especially loop quantum cosmology
\cite{LivRev,Springer} (see \cite{SIGMA} for a recent review), begin with a
similar but reversed setting, now dealing with discrete space and almost
periodic or compactified momentum space. In this case, space is not
represented by position coordinates but by geometrical quantities such as the
total volume $V$ of an isotropic universe model, or in general by points in
minisuperspace. The momentum $P$ is then related to curvature components or,
in cosmology, the Hubble parameter. As with Bloch states, the Hilbert space
(in the momentum representation) is spanned by states
\begin{equation} \label{Pstates}
 \psi_n^{(\epsilon)}(P)= \exp(i\mu_n^{(\epsilon)}P)
\end{equation}
with the same form (\ref{mu}) of $\mu_n^{(\epsilon)}$ as before, except that
$q_0$ is to be replaced by a quantity $P_0$ signaling the periodicity of $P$
\cite{IsoCosmo,Bohr}.\footnote{Another difference (but one not relevant here)
  is that $\bigoplus_{\epsilon}{\cal H}_{\epsilon}$ is taken as the full
  Hilbert space of the model, with all states (\ref{Pstates})
  normalizable. One can view these wave functions as being supported on the
  Bohr compactification of the real line \cite{Bohr}, with an invariant
  measure that differs from the one usually used in quantum mechanics. At this
  level, one is dealing with a non-separable Hilbert space, but superselection
  is often introduced at the dynamical level. Then, one can again assume fixed
  $\epsilon$ without loss of generality.} These are the main aspects of loop
quantum cosmology we need in this article; see Appendix~\ref{a:LQC} for more
details.

In addition to technical properties of the dynamics, there is a key
physical difference between the treatment of Bloch waves as a model
for condensed matter physics and isotropic loop quantum cosmology as a
model for quantum gravity: Bloch states represent a system in which
the position coordinate $q$ is almost periodic, and thus its momentum
is discrete. The regime of distances $q\gg q_0$ {\em much larger} than
the periodicity is easily accessible by classical physics, and one is
interested in uncovering what happens at smaller distances near the
scale of periodicity. In loop quantum cosmology, on the other hand,
the (momentum-like) expansion rate $P$ is almost periodic while the
size $V$ is discrete. Moreover, it is the {\em low-curvature} regime
$P\ll P_0$ which is easily accessible by classical physics and one is
interested in uncovering what happens at large curvature near
$P_0$. This point plays an important role regarding the specific
questions one tries to address. In this article, we will mainly be
concerned with the quantum-cosmology-like situation, probing the
quantum system well below the periodicity scale.  This regime will be
implemented by the approximations used.

\subsection{Uncertainty with periodic momenta}
\label{s:Periodic}

Motivated by the examples of discrete systems, we assume a general
class of models with a periodicity condition on the momentum: wave
functions $\phi(p)$ in momentum space obey $\phi(-p_0/2)=\phi(p_0/2)$
for some momentum value $p_0$. Compared to the more general discussion
before, we set $\epsilon=0$ without loss of generality; non-zero
values will simply shift the lattice structure we obtain in position
space. Here, the superselection assumption is important. The conjugate
variable $q$ is then quantized to an operator with discrete spectrum
$q_n=2\pi\hbar n/p_0$ with integer $n$. We will analyze the possible
values of uncertainties that can be realized in the set ${\cal
F}_{\bar{q}}$ of wave functions that possess some fixed position
expectation value $\bar{q}=\langle\hat{q}\rangle$. In particular, we
ask how small the position fluctuation $\Delta q$ can be in this set,
or how well we can localize a particle at position $\bar{q}$. Our aim
is to derive a function $\Delta q_{\rm min}(\bar{q})$ which determines
the minimally possible uncertainty for localization at $\bar{q}$.

If we choose $\bar{q}$ to be one of the lattice points, $q_n$, we may
localize the particle arbitrarily sharply because we could choose the
state to be the $\hat{q}$-eigenstate with eigenvalue $q_n$. Thus,
$\Delta q_{\rm min}(q_n)=0$. As we will show now, for all other values
of $\bar{q}$ the minimum uncertainty is not zero.

Without loss of generality, we then choose
$q_0=0<\bar{q}<q_1=2\pi\hbar/p_0$. A corresponding wave function can no longer
be a position eigenstate, and in order to achieve minimum position uncertainty
we should choose a superposition of the eigenstates with position eigenvalue
zero and $q_1$:
\[
 \phi_{\bar{q}}(p)= a e^{-i q_0p}+ be^{-i q_1p}= a+b e^{-2\pi i
 p/p_0}\,.
\]
With normalization, $|a|^2+|b|^2=1/p_0$. Moreover, we
straightforwardly compute
\begin{equation}
 \bar{q}=2\pi\hbar|b|^2  \quad,\quad \langle \hat{q}^2\rangle= 4\pi^2\hbar^2
 |b|^2/p_0\,.
\end{equation}
Eliminating $|b|$, we obtain 
\begin{equation} \label{Deltaqmin}
 \Delta q_{\rm min}(\bar{q})= \sqrt{\bar{q}(q_1-\bar{q})}
 \quad\mbox{for}\quad 0\leq \bar{q}\leq q_1 \,,
\end{equation}
extended periodically over the whole $q$-axis, consistent with the findings in
\cite{UncertSpectrum}.  For sectors with $\epsilon\not=0$, we obtain the same
formula just with $q_1$ interpreted as the lattice spacing
$L=q_{1+\epsilon}-q_{\epsilon}=q_1$.  The minimal uncertainty indeed vanishes
for $\bar{q}$ a lattice point, and is at most half the lattice spacing:
$\Delta q_{\rm min}\leq L/2$. At this stage we see the importance of the
superselection assumption.  Without it, we could have made the minimal
uncertainty arbitrarily small for all $\bar{q}$; for every $\bar{q}$, there is
an $\epsilon$-sector containing a $\hat{q}$-eigenstate with eigenvalue
$\bar{q}$. From the perspective of minimally possible position uncertainty,
the discreteness is thus noticeable only if the $\epsilon$-sector is fixed,
for instance derived from other observations. On one hand, if all
$\epsilon$-sectors were allowed, we could localize at every point with
absolute precision.  On the other hand, if instead in momentum space the
boundary condition of periodicity up to a phase $e^{i\epsilon}$ is replaced by
Dirichlet boundary conditions then $\hat{p}$ is symmetric but it is not
self-adjoint. In this case, at no point could the position be resolved to
absolute precision, leading to a global finite $\Delta q_{\rm min}$. We will
also encounter this case below.

We now turn to momentum uncertainties. The minimum position
uncertainty can be used to probe the lattice structure only if the
resolution of our measurements is close to the lattice
spacing. Moreover, the $\epsilon$-sector would have to be determined
by independent means. An important question then is how the lattice
structure can be noticed if measurements are done at energies which
may be high, but not high enough to resolve the lattice.  One way that 
may offer an opportunity to overcome this problem may be to test
 for small deviations from the usual uncertainty relations, namely by checking the 
relationship between both position and momentum fluctuations. Before we enter a
more detailed discussion of generalized uncertainty relations, for later comparisons it
will be useful to continue with the question of localization and
compute some of the corresponding momentum uncertainties. \rm

Again, we choose a position eigenstate of one of the lattice points,
without loss of generality at $\bar{q}=q_0=0$. Then, $\phi_{q_0}(p)=
1/\sqrt{p_0}$. In addition to $\bar{q}=0$ and $\Delta q=0$, we have
$\bar{p}=\langle\hat{p}\rangle=0$ and $\Delta p=p_0/2\sqrt{3}$. One of
the consequences of discreteness is that $\Delta q=0$ is possible with
finite $\Delta p$, clearly requiring modified uncertainty relations
compared to the continuum case. It will also be useful to consider
higher moments of the state, { in particular 
\begin{equation}
 \Delta(p^n):=
  \langle(\hat{p}-\bar{p})^n\rangle= \frac{p_0^n}{2^n(n+1)}
\end{equation}
for even $n$ while $\Delta(p^n)=0$ if $n$ is odd.  The series
  $\Delta(p^n)/p_0^n$ thus falls off for increasing $n$.}

\subsection{Generalized uncertainty relations}

As the preceding example demonstrates, quantum systems with discrete or
periodic structures in phase space cannot obey the usual uncertainty relation
$\Delta q\Delta p\geq \hbar/2$ of quantum mechanics because the lattice
structure makes it possible for $\Delta q$ to vanish at finite $\Delta
p$. Nevertheless, we still expect some form of uncertainty relation to apply;
after all, at distance scales much larger than the lattice spacing we should
be able to recover standard continuum quantum mechanics.  A common way to
parameterize generalized uncertainty relations is
\begin{equation} \label{GUP}
 \Delta q\Delta p\geq \frac{\hbar}{2} \left(1+\alpha(\Delta p)^2
 +\beta(\Delta q)^2+\gamma\right)\,,
\end{equation}
considered first in \cite{UncertQuantGroup}, see also, e.g., 
\cite{ShortDistance,ShortDistanceThree,NonLocalGUP}.

The parameters $\alpha,\beta$ and $\gamma$ are independent of $\Delta q$ and
$\Delta p$ but in general may depend on expectation values of the overall
state. Dimensional analysis of the correction terms in Eq.~(\ref{GUP})
indicates that these parameters are not purely quantum corrections, as perhaps
motivated by quantum gravity. If one uses only Planck's constant and the
Planck length, dimensionally we must have $\alpha\propto \ell_{\rm
  P}^2/\hbar^2=G/\hbar$ and $\beta\propto 1/\ell_{\rm P}^2= 1/G\hbar$, both
proportional to $\hbar^{-1}$. As quantum corrections, this behavior is
unsuitable because the terms $G(\Delta p)^2/\hbar$ and $(\Delta q)^2/G\hbar$
do not necessarily go to zero for $\hbar\to0$, with semiclassical fluctuations
squared usually being about the size of $\hbar$. Generalized uncertainty
principles thus require either modifications to the quantum algebra of basic
operators and even the classical symplectic structure, or an additional scale
not directly related to $\hbar$. This additional scale could be the
band\-limit of a fundmental bandlimitation \cite{UncertBandlimit}, the size of
fundamental extended objects, or the periodicity or discreteness scale
considered in this paper.

\subsubsection{Implications}

In Eq.~(\ref{GUP}), let us first consider the case where
$\alpha,\beta>0$, $\gamma>-1$. If also $\alpha\beta \ge 1/\hbar^2$, then this
uncertainty principle has no solutions, i.e., we can rule out this
case: {for $x:=\Delta q/\sqrt{\alpha}\hbar$ and
$y:=\sqrt{\alpha}\Delta p$ the relation implies the impossible
relationship $(x-y)^2\leq-(1+\gamma)<0$.}  Else, if
$\alpha,\beta>0$ and $\alpha\beta \le 1/\hbar^2$ , then the
uncertainty relation (\ref{GUP}) arises from the commutation relation
\begin{equation} \label{ModComm}
 [\hat{q},\hat{p}]= i\hbar (1 + \alpha \hat{p}^2 + \beta \hat{q}^2).
\end{equation}
through $\Delta A \Delta B \ge \frac{1}{2} \vert\langle [A,B]\rangle\vert$
which holds for any symmetric or self-adjoint operators $A,B$ on any domain on
which they and their commutator can act. Notice that (\ref{ModComm}) induces
an uncertainty relation of the type of (\ref{GUP}) with a generally
non-vanishing $\gamma$ that depends on $\langle\hat{q}\rangle$ and
$\langle\hat{p}\rangle$.  A Hilbert space representation can be constructed
using $\rho$-deformed raising and lowering operators, $\hat{a}$,
$\hat{a}^\dagger$. (In the literature on quantum groups, the parameter $\rho$
is usually denoted $q$, but we here use the symbol $q$ for the position
operator).  Namely, in this case the operators $\hat{q}$ and $\hat{p}$ can be
represented through
\begin{equation}
\hat{q} := \frac{1}{\sqrt{2\beta(1/\hbar\sqrt{\alpha\beta}-1)}}~
(\hat{a}^\dagger +\hat{a}) \label{posq}
\end{equation}
\begin{equation}
\hat{p}:= \frac{i}{\sqrt{2\alpha(1/\hbar\sqrt{\alpha\beta}-1)}}~
(\hat{a}^\dagger-\hat{a})  \label{momp}
\end{equation}
where $\hat{a}$, $\hat{a}^\dagger$ obey
\begin{equation}
\hat{a} \hat{a}^\dagger - \rho \hat{a}^\dagger \hat{a}=1
\end{equation}
with 
\begin{equation}
\rho := \frac{1+\hbar\sqrt{\alpha\beta}}{1-\hbar\sqrt{\alpha\beta}}
\end{equation}
Note that $\rho \in (1,\infty)$. As usual, the Hilbert space together with a representation of $\hat{q}$ and $\hat{p}$ can be
  constructed by the Fock method on a state $\vert 0\rangle$ obeying
  $\hat{a}\vert 0\rangle=0$.

For $\beta=0$, the
representations of the generalized commutation relation
$[\hat{q},\hat{p}]= i\hbar(1+\alpha\hat{p}^2)$ are discussed in
\cite{NonLocalGUP}, where it was found that their properties qualitatively depend on
the sign of $\alpha$: 
\begin{itemize}
\item For $\alpha<0$, there
are finite-dimensional representations. In infinite-dimensional ones,
$\hat{p}$ is a bounded operator and has a finite range of
eigenvalues; $\hat{q}$ possesses self-adjoint extensions whose
spectra are continuous.
\item For $\alpha>0$, $\hat{p}$ has a continuous spectrum comprised of
the entire real line. The self-adjoint extensions of $\hat{q}$  possess
discrete parts to their spectra and normalizable eigenvectors.
\end{itemize}

Let us now return to Eq.~(\ref{GUP}) for generic $\alpha,\beta$ and
$\gamma$. It is of particular interest to probe the smallest allowed
scales by determining how small $\Delta q$ can be made. In the case
$\alpha>0$, $\beta>0$, $\gamma>-1$, $\alpha\beta\leq 1/\hbar^2$ of
above, it is known that $\Delta q$ possesses a non-vanishing minimum
overall, as we will recover as a special case. But we also expect
that, in other cases, the vanishing of $\Delta q$ may be possible for
finite $\Delta p$ as required for lattice models.

We begin by noticing that saturating the
uncertainty relation requires
\begin{equation}
 \Delta q = \frac{\Delta p\pm \sqrt{(1-\hbar^2\alpha\beta)(\Delta
 p)^2- \hbar^2 \beta (1+\gamma)}}{\hbar \beta}\,.
\end{equation}
For fixed $\alpha$, $\beta$ and $\gamma$ this expression is minimized
for
\[
 (\Delta p)^2 = \frac{1+\gamma}{\alpha(1-\hbar^2\alpha\beta)}
\]
such that the uncertainty in position is bounded from below by
\begin{equation} \label{Deltaq}
 \Delta q = \hbar \sqrt{\frac{\alpha (1+\gamma)}{1-\hbar^2\alpha\beta}}
\end{equation}
provided the square root is well-defined.  For $\alpha(1+\gamma)>0$ a
positive lower bound for the position uncertainty results
independently of the momentum uncertainty as in the example of Section
\ref{s:Periodic} in the case of Dirichlet boundary conditions. If
instead $\alpha<0$ and $1+\gamma>0$, then the generalized uncertainty
relation Eq.~(\ref{GUP}) allows $\Delta q$ to vanish at finite $\Delta
p=\sqrt{-(1+\gamma)/\alpha}$, qualitatively similar to our example
above when fixing an $\epsilon$-sector. This confirms our
expectation that the coefficients in generalized uncertainty
relations, and especially their signs, carry information about
underlying discrete structures.

Indeed, even if no direct information is available about the boundary
conditions in momentum space, such as the specific $\epsilon$-sector,
indications of negative values of $\alpha$ (for positive $1+\gamma$)
would imply agreement with the discrete model, while positive $\alpha$
would correspond to a finite lower bound to the position uncertainty
(\ref{Deltaq}).  

\subsubsection{Representations}

Properties of operators and Hilbert-space representations can be surprisingly
subtle in the context of generalized uncertainty relations. In order to
illustrate this, let us have a closer look at the case of $\alpha, \beta >0$,
i.e., at the case of a finite lower bound to the position uncertainty. The
operators $\hat{q}$ and $\hat{p}$ then act via Eqs.~(\ref{posq}),
(\ref{momp}), on the domain, $D$, of all finite complex linear combinations of
the basis vectors $(a^\dagger)^n\vert 0 \rangle$. Clearly, $D$ is dense in the
Hilbert space, $\cal{H}$, of all (finite or infinite) normalizable linear
combinations of the vectors $(a^\dagger)^n\vert 0 \rangle$. It is
straightforward to verify that the commutation relation holds on $D$ and that
$\hat{q}$ and $\hat{p}$ are symmetric operators, i.e., that 
all their expectation values are real: $\langle \phi \vert
\hat{q}\vert \phi\rangle\in \mathbb{R}$ and $\langle \phi \vert \hat{q}\vert
\phi\rangle\in \mathbb{R}$ for all $\vert\phi\rangle \in D$. As always in
quantum mechanics, we obtain the physical domain $D_{\rm physical}$ by
enlarging $D$ so as to include as many infinite linear combinations of the
basis vectors $(a^\dagger)^n\vert 0 \rangle$ as possible. Concretely, $D_{\rm
  physical}\subset\cal{H}$ is the maximal domain on which the commutation
relation holds. This means that $D_{\rm physical}$ is the maximal domain on
which the images of all operators that occur in the commutation relations are
contained in the Hilbert space. Therefore, $D_{\rm physical}$ is the set of
all $\vert \phi\rangle \in \cal{H}$ for which $\hat{q}\vert\phi\rangle\in
\cal{H}$, $\hat{p}\vert\phi\rangle\in \cal{H}$,
$\hat{q}\hat{p}\vert\phi\rangle\in \cal{H}$,
$\hat{p}\hat{q}\vert\phi\rangle\in \cal{H}$, $\hat{q}^2\vert\phi\rangle\in
\cal{H}$ and $\hat{p}^2\vert\phi\rangle\in \cal{H}$.

In this context, let us recall that the presence of finite lower bounds to
$\Delta q$ and $\Delta p$ precludes the existence of eigenvectors of $\hat{q}$
or $\hat{p}$ in $D_{\rm physical}$ since they would have vanishing variance,
$\Delta q=0$ or $\Delta p=0$. The lower bounds even preclude the existence of
sequences of physical vectors whose variance, say $\Delta q$, goes to zero
(even while allowing that $\Delta p$ might diverge). As one might expect,
therefore, $\hat{q}$ and $\hat{p}$ on $D_{\rm physical}$ have no complete
  spectral decomposition and therefore cannot be 
self-adjoint \cite{UncertQuantGroup}. The phenomenon that operators, such as
$\hat{q}$ and $\hat{p}$, are symmetric on a domain, here $D_{\rm physical}$,
without being self-adjoint, is a subtlety that can occur only in
infinite-dimensional Hilbert spaces.

Interestingly, the detailed functional analysis of these operators
shows that $\hat q$ and $\hat{p}$ individually do possess extensions
of their domain on which they become self-adjoint.  In particular,
there exists a family of enlarged domains $D_{q,\alpha}$, parametrized
by $\alpha \in [0,1)$, obeying $D_{\rm physical}\subset
D_{q,\alpha}\subset\cal{H}$ such that for each fixed $\alpha$ the
extended $\hat{q}_\alpha$ which acts on $D_{q,\alpha}$ is self-adjoint
and has a discrete spectrum, $\{q_{n,\alpha}\}_{n\in\mathbb{Z}}$,
along with normalizable eigenvectors $\{\vert
q_{n,\alpha}\rangle\}_{n\in\mathbb{Z}}$. It has been shown that as
$\alpha$ runs through the interval $[0,1)$, the corresponding discrete
grids of eigenvalues $\{q_{n,\alpha}\}$ cover the real line exactly
once,
$\bigcup_{\alpha\in[0,1)}\{q_{n,\alpha}\}_{n\in\mathbb{Z}}=\mathbb{R}$. The
fact that $\hat{q}_\alpha$ possesses eigenvectors $\{\vert
q_{n,\alpha}\rangle\}_{n\in\mathbb{Z}}$, for which $\Delta q_\alpha
=0$, is consistent with the fact that we have a positive lower bound
(\ref{Deltaq}) for $\Delta q$.  The reason is of course that the
eigenvectors $\vert q_{n,\alpha}\rangle$ are in $D_{q,\alpha}$ but not
in $D_{\rm physical}$.

Nevertheless, while keeping in mind that the vectors $\vert
q_{n,\alpha}\rangle$ are not in the physical domain, we may of course
utilize the fact that any such set of eigenvectors, $\{\vert
q_{n,\alpha}\rangle\}_{n\in\mathbb{Z}}$, for any fixed $\alpha$, is a
basis in the Hilbert space. Namely, we can use the fact that any
physical state $\vert \phi\rangle\in D_{\rm physical}$ is completely
specified by its coefficients $\langle q_{n,\alpha}\vert\phi\rangle$
in the Hilbert basis $\{\vert
q_{n,\alpha}\rangle\}_{n\in\mathbb{Z}}$. This means that all physical
kinematics and dynamics, i.e., that all relationships and maps between
vectors in $D_{\rm physical}$ can be described as relationships and maps
between the coefficients of these vectors in the basis $\{\vert
q_{n,\alpha}\rangle\}_{n\in\mathbb{Z}}$. The theory can therefore be
viewed as a theory living on the discrete set of positions
$\{q_{n,\alpha}\}_{n\in\mathbb{Z}}$ for some fixed
$\alpha$. Nevertheless, this is not a discrete theory in the usual
sense because the discretization is optional and one may freely change
to describing the same physical dynamics and kinematics on any other
grid of positions $\{q_{n,\alpha'}\}$ for some other $\alpha'$. This
equivalence of a whole family of discrete representations of a theory
is made possible by the fact that the finite lower bound $\Delta
q_{\rm min}$ makes these discretizations physically indistinguishable by
any physical fields $\vert \phi\rangle \in D_{\rm physical}$.

This mathematical structure provides a generalization of Shannon
sampling theory, see \cite{UncertBandlimit-prl}, with $\Delta q_{\rm min}$
playing the role of a finite bandwidth. (Shannon sampling
theory provides the link between discrete and continuous
representations of information and it is used ubiquitously in signal
processing and communication engineering.) The case $\alpha >0$
therefore describes a space which is simultaneously discrete and
continuous in the same way that information can be continuous and
discrete, see \cite{UncertBandlimit}.

\subsubsection{Back to generalized uncertainty relations}

Our interest now will be to understand the interplay between lower
bounds to position uncertainties and actual spatial discreteness in a
way that is independent of representations and their functional
analytic subtleties.

To analyze the relationship between a discrete length and coefficients
in a generalized uncertainty principle, we here take a route on which
we start with a conventional quantization of a fundamentally discrete
quantum system. From this, we derive a generalized uncertainty
principle of the form (\ref{GUP}), with uniquely determined
coefficients. Our methods will be representation-independent, thus
avoiding the need to address questions of super\-selection or
domains. Although the example we study is simple, it should be able to
serve as a model for analogous derivations to be performed if one
wants to derive predictions for low-energy effects of fundamentally
discrete systems, such as some versions of quantum gravity.

\section{Quantum mechanics on a circle}
\label{s:Circle}

In order to study the effects of the discreteness of the
position, $q$, perturbatively, we will now use a simple system 
given by a quantized phase space of a cylinder
where momentum $p$ has periodicity $p_0$, and derive uncertainty
relations in an expansion by $p/p_0$. According to the
discussion above, this is the regime of interest in quantum
cosmology. The expansion can be done in a systematic and
representation-independent way by computing higher moments of a state,
and it provides specific coefficients which one can compare with the
general form (\ref{GUP}). Our techniques are motivated by a general
scheme of effective equations in a canonical setting, which was
developed in \cite{EffAc,EffectiveEOM,Karpacz}. Such equations have
been derived in loop quantum cosmology \cite{BouncePert}, for which
the circle system provides a model capturing the characteristic
representation. In fact, quantum mechanics on a circle can be seen as
a sector in the Hilbert space of loop quantum cosmology, just as the
set of all Bloch states is split into sectors of functions periodic up
to phase. Being based on the same techniques, generalized uncertainty
relations and effective equations may thus be combined for further
phenomenological applications of quantum cosmology.

We present a brief overview of this simple well-known system in order to
introduce our notation. Classical variables are a canonical pair $(q,p)$ with
Poisson bracket $\{q,p\}=1$. In analogy with loop quantum cosmology we choose
the momentum $p$ to be periodic, such that $p$ is the angle of a circle and
thus takes values in $S^1$. Then, $q$ becomes discrete upon quantization. The
phase space can be described by a complete set of phase space variables
$(q,\sin(2\pi p/p_0),\cos(2\pi p/p_0))$ where $p_0$ is the periodicity of $p$
which, $p$ being a dimensionless angle, can be fixed to $p_0=1$ but will be
more useful for future expansions if kept unspecified. Instead of using the
sine and cosine, it is more convenient to use the complex-valued function
$h:=\exp(2\pi ip/p_0)$ and its complex conjugate $h^*$, subject to the reality
condition $h^*h=1$. These basic functions satisfy the non-canonical algebra
\begin{equation} \label{qh}
 \{q,h\} = \frac{2\pi i}{p_0} h \quad,\quad \{q,h^*\} = 
-\frac{2\pi i}{p_0}h^*
 \quad,\quad \{h,h^*\}=0
\end{equation}
under taking Poisson brackets.

The quantum theory can be formulated on the Hilbert space $L^2(S^1,{\rm d}
p/p_0)$ which has an orthonormal basis $\{\vert n\rangle\}_{n\in{\mathbb N}}$,
with momentum representation $\langle p|n\rangle = \exp(2\pi inp/p_0)$.  The
variable $q$ is directly quantized to become a multiplication operator acting
by $\hat{q}|n\rangle = 2\pi \hbar p_0^{-1}n|n\rangle$ which shows the
discreteness of its spectrum. As before, wave functions need not be strictly
periodic but could also be chosen periodic up to a phase:
$\psi(p+p_0)=\exp(i\epsilon)\psi(p)$ with $\epsilon\in{\mathbb R}$. This is
sufficient to ensure that the probability density is single-valued on the
circle, and introduces a 1-parameter family of inequivalent representations
for $\epsilon\in [0,2\pi)$. They are inequivalent because the
$\hat{q}$-spectrum possesses the eigenvalues $2\pi\hbar(n+\epsilon)/p_0$ which
depend on $\epsilon$.  (We remark that we are now dealing with a closed circle
instead of an interval with boundary conditions, so that non-strict
periodicity may seem impossible to impose. Nevertheless, the corresponding
Hilbert spaces can be formulated as function spaces on non-trivial line bundle
over the circle, but we will not explicitly require these structures here.)
There is no operator for $p$, however, because as a multiplication operator it
would not map a basis state into another allowed state. Another way to see
that such an operator cannot exist is to note that it would generate
infinitesimal translations in $q$, which is not possible due to the
discreteness of the $\hat{q}$-spectrum. There are, instead, well-defined
operators for our basic functions $h$ and $h^*$, satisfying $\hat{h}|n\rangle
= |n+1\rangle$ and $\widehat{h^*}|n\rangle= |n-1\rangle$. The reality
condition for $p$ is satisfied since $\hat{h}\widehat{h^*}=\hat{1}$ and
$\widehat{h^*}=\hat{h}^{\dagger}$.

\subsection{Moment algebra}

Irrespective of the representation chosen, these basic operators satisfy
the commutator algebra
\begin{equation} \label{algebra}
 [\hat{q},\hat{h}]= -\frac{2\pi \hbar}{p_0}\hat{h} \quad,\quad
 [\hat{q},\hat{h}^{\dagger}]= \frac{2\pi\hbar}{p_0}\hat{h}^{\dagger}
 \quad,\quad [\hat{h},\hat{h}^{\dagger}]=0
\end{equation}
which faithfully quantizes the classical basic algebra. The following
calculations and our main results will make use only of this algebra
and the reality condition, as well as the general Schwarz inequality;
therefore they will be manifestly representation-independent.

Instead of working with wave functions as states, we will be using
only the algebra (\ref{algebra}) and functionals on it, suggestively
characterized by expectation values $q=\langle\hat{q}\rangle$,
$h=\langle\hat{h}\rangle$, $h^*=\langle\hat{h}^{\dagger}\rangle$ and
moments
\begin{equation}
 \Delta(q^ah^b):=\left\langle \left((\hat{q}-q)^{a} (\hat{h}-h)^b\right)_{\rm
 Weyl}\right\rangle
\end{equation}
of expectation values in Weyl ordering, where $a,b\in\mathbb{N}$ and $a+b\geq
2$.  These variables form an (over-) complete set of functionals
assigning complex numbers to the operators in our algebra. It follows
  from Hamburger's theorem that the probability density of a wave function can
  be reconstructed from the moments $\Delta(q^n)$, while the phase of the wave
  function can be found using moments involving $h$. For a pure state, the set
  of all moments is overcomplete. The additional freedom in the set of moments
  allows one to include mixed states as well.)  The moments can be varied
independently of expectation values to describe different states, provided
they respect inequalities and reality conditions as discussed below. They are
also useful for an analysis of coherent-state properties as e.g.\ in
\cite{BounceCohStates}, which provides a link to the uncertainty
relation. Our analysis here provides an independent and more
  direct relationship.  From now on, we denote expectation values of basic
operators by $q$ and $h$ without distinguishing them from the classical
variables. This convention simplifies the notation and should not give rise to
confusion.

Often, it is more convenient to work directly with equations for the moments
rather than taking the detour of wave functions or density matrices,
presenting a complete description from a more algebraic and
representation-independent viewpoint. All crucial aspects of the system are
then contained in the basic algebra, which in our case in particular means to
use $\hat{h}$ as a basic operator on the circle, possibly combined with a
Hamiltonian or a constraint. The main challenge then is to organize the
infinitely many variables provided by the moments, and the equations of motion
they must fulfill. An example where these equations can be organized in
manageable ways is given by semiclassical regimes, in which moments of high
order are small, but the treatment is not restricted to this case.  Our
approximation below will only assume the momentum (related to $h$) to be small
compared to $p_0$, and any moments involving $p$ (relative to $p_0$) to fall
off with increasing order as they do for semiclassical states but not only for
such states; with these assumptions, fluctuations may still be large.
Moreover, the size of the $q$-moments will remain unrestricted and need not be
small compared to powers of $\hbar$.  An advantage of the use of expectation
values and moments instead of wave functions is not only the representation
independence but also its larger generality: it includes mixed states as well
as pure ones.

We will be working mainly with moments of lower order where $a+b$ is
small. For better clarity, we will then replace the superscript
``$a,b$'' by a list of operators used in the moments. For instance, we
have the $h$-variance $\Delta (h^2)\equiv(\Delta h)^2=:\Delta h^2$
and the covariance
\[
 \Delta(qh)= \frac{1}{2} \langle (\hat{q}-q)(\hat{h}-h)+
 (\hat{h}-h)(\hat{q}-q)\rangle= \frac{1}{2}
 \langle\hat{q}\hat{h}+\hat{h}\hat{q}\rangle- qh\,.
\]

\subsection{Reality conditions}

Expectation values and second-order moments are related to one another
by the reality condition: taking an expectation value of the relation
$\hat{h}\hat{h}^{\dagger}=\hat{1}$ implies
\begin{equation} \label{RealityI}
 hh^*= 1-\Delta(hh^*)\,.
\end{equation}
This relation can be interpreted as reducing the number of independent
expectation values of the basic variables to the canonical value two,
such as $q$ and ${\rm Re(}h)$ (at fixed moments).

Similarly, at higher orders of the moments, we obtain additional reality
conditions which reduce the number of moments to the canonical values as
  already used in \cite{HighDens}. For the second-order moments, we begin
with the identities $\hat{h}^2\hat{h}^{\dagger}= \hat{h}$ and
$\hat{q}\hat{h}\hat{h}^{\dagger}=\hat{q}$ that follow from
$\hat{h}\hat{h}^{\dagger}=\hat{1}$, and take expectation values. With some
symmetric reorderings according to the definition of the moments, we obtain
\begin{eqnarray}
 h^*\Delta h^2+h\Delta(hh^*)&=& -\Delta(h^2h^*)\label{RealityII}\\
 h^*\Delta(qh)+h\Delta(qh^*)&=& -\Delta(qhh^*)\label{RealityIII}\,.
\end{eqnarray}
The first equation is complex and implies two independent conditions
for the moments, while the second equation is real. There are thus
three conditions to restrict the second-order moments (at fixed
third-order ones) to the correct canonical number: out of six initial
moments $\Delta q^2$, ${\rm Re}\Delta(qh)$, ${\rm Im}\Delta(qh)$,
$\Delta(hh^*)$, ${\rm Re}\Delta h^2$ and ${\rm Im}\Delta h^2$, three
moments are left independent, amounting to two fluctuations and one
correlation.

\subsection{Uncertainty relations}

The main interest here lies in uncertainty relations which can be
formulated in terms of the moments even if they are not used for a
canonical pair $(q,p)$ but for a pair of our basic operators. (See
e.g.\ \cite{BounceCohStates} for more details.) As usual, from the
Schwarz inequality one derives
\begin{equation}
 \Delta A^2\Delta B^2- \Delta(AB)^2 \geq \frac{1}{4}\langle
 i[\hat{A},\hat{B}]\rangle^2
\end{equation}
for any pair $(\hat{A},\hat{B})$ of self-adjoint or symmetric operators. In our
case, we can form three pairs of self-adjoint operators from the set
$(\hat{q},\hat{h}+\hat{h}^{\dagger}, i(\hat{h}-\hat{h}^{\dagger}))$,
giving uncertainty relations
\begin{eqnarray} \label{UncertI}
&& \Delta q^2\Delta(h+h^*)^2- \Delta(q(h+h^*))^2 = 2\Delta q^2({\rm
   Re}\Delta h^2+\Delta(hh^*))- 4({\rm Re}\Delta(qh))^2\nonumber\\
&&\geq -\frac{\pi^2\hbar^2}{p_0^2} (h-h^*)^2
\end{eqnarray}
for $\hat{A}=\hat{q}$ and $\hat{B}=\hat{h}+\hat{h}^{\dagger}$,
\begin{eqnarray} \label{UncertII}
&&\Delta q^2\Delta(i(h-h^*))^2- \Delta(qi(h-h^*))^2 =
   2\Delta q^2(-{\rm Re}\Delta h^2+\Delta(hh^*))- 4({\rm Im}\Delta(qh))^2
\nonumber\\
 &&\geq \frac{\pi^2\hbar^2}{p_0^2} (h+h^*)^2
\end{eqnarray}
for $\hat{A}=\hat{q}$ and $\hat{B}=i(\hat{h}-\hat{h}^{\dagger})$, and
\begin{eqnarray}\label{UncertIII}
&& \Delta(h+h^*)^2\Delta(i(h-h^*))^2
 -\Delta((h+h^*)i(h-h^*))^2\nonumber\\
&& = 4\left(\Delta(hh^*)^2-({\rm
   Re}\Delta h^2)^2\right)- 4({\rm Im}\Delta h^2)^2\geq 0
\end{eqnarray}
for $\hat{A}=\hat{h}+\hat{h}^{\dagger}$ and
$\hat{B}=i(\hat{h}-\hat{h}^{\dagger})$.

In semiclassical regimes, with moments of third or higher orders ignored, one
can use the reality conditions to show that (\ref{UncertII}) implies
(\ref{UncertI}) and (\ref{UncertIII}). If moments of higher order are kept,
(\ref{UncertI}) and (\ref{UncertIII}) in combination with (\ref{UncertII}) and
the reality conditions imply conditions for third-order moments, an
  example for higher-order uncertainty relations. For instance,
(\ref{RealityII}), solved for $h^*\Delta h^2$ and then taken in its absolute
value, implies
\[
 |h|^2\left(\Delta(hh^*)^2-|\Delta h^2|^2\right)= -|\Delta(h^2h^*)|^2- 2{\rm
  Re}(h^*\Delta(hh^*)\Delta(h^2h^*))
\]
and then
\begin{equation} \label{RealityIV}
 -2{\rm Re}(h^*\Delta(hh^*)\Delta(h^2h^*))\geq |\Delta(h^2h^*)|^2
\end{equation}
with (\ref{UncertIII}).

Given that $\frac{i}{2}(\hat{h}-\hat{h}^{\dagger})$ corresponds to the sine of
$\hat{p}$, which should reduce to $\hat{p}$ when acting on states
  supported only on small $p$, we expect that it is (\ref{UncertII}) which
reduces to the standard uncertainty relation when $p$ is small enough so that
the periodicity can be ignored. To confirm this expectation, we first consider
only leading orders in the $p_0^{-1}$-expansion: we expand the operator
\begin{equation}
 \hat{h}= 1+\frac{2\pi i}{p_0}\hat{p}- \frac{2\pi^2}{p_0^2}\hat{p}^2
+\cdots,
\end{equation}
which is valid on a set of states supported on values of $p$ small
compared to $p_0$, and then compute the moments for the expansion.
To leading order in $p_0^{-1}$, we need only the term $\hat{h}-h=2\pi
ip_0^{-1} (\hat{p}-p)+\cdots$, for which
\begin{equation}
 \Delta(hh^*)= \langle(\hat{h}-h)(\hat{h}^{\dagger}-h^*)\rangle=
 \frac{4\pi^2}{p_0^2}\Delta p^2  +\cdots
\end{equation}
and
\begin{equation}
 \Delta h^2= \langle(\hat{h}-h)^2\rangle=
 -\frac{4\pi^2}{p_0^2}\Delta p^2  +\cdots \,.
\end{equation}
(As one can easily verify to this order, the reality condition
$\Delta(hh^*)=1-|h|^2$ is identically satisfied in terms of the $p$-moments.)

For mixed moments we have to be more careful with the ordering:
\begin{equation}
 \Delta(qh)=\frac{1}{2}
 \langle\hat{q}\hat{h}+\hat{h}\hat{q}\rangle-qh
 = \frac{i\pi}{p_0} \langle\hat{q}\hat{p}+\hat{p}\hat{q}\rangle
 -\frac{2\pi i}{p_0}qp+\cdots = \frac{2\pi i}{p_0}\Delta(qp)+\cdots\,.
\end{equation}
Inserting this in (\ref{UncertII}) provides the uncertainty product
\begin{equation}
 2\Delta q^2(\Delta(hh^*)- {\rm Re}\Delta h^2)- 4({\rm Im}\Delta(qh))^2=
 \frac{16\pi^2}{p_0^2} (\Delta q^2\Delta p^2-\Delta(qp)^2)+\cdots
\end{equation}
which together with
\[
 \frac{\pi^2\hbar^2}{p_0^2}(h+h^*)^2= \frac{4\pi^2\hbar^2}{p_0^2} +\cdots
\]
results in the standard uncertainty relation
\begin{equation}
 \Delta q^2\Delta p^2- \Delta(qp)^2\geq \frac{\hbar^2}{4}\,.
\end{equation}
Equations (\ref{UncertI}) and (\ref{UncertIII}) are satisfied identically
to this order up to $p_0^{-2}$.

\subsection{Corrections to the uncertainty relation}

Corrections do arise, however, if we expand to higher orders in
$p_0^{-1}$, in which case we will obtain a generalized uncertainty
relation as we demonstrate now. For instance, expanding to the next
order on the right-hand side of the uncertainty relation
(\ref{UncertII}) gives
\begin{equation}
 \frac{1}{2}(h+h^*)= 1-\frac{2\pi^2}{p_0^2}\left(p^2+\Delta p^2\right)
+\cdots\,.
\end{equation}
These corrections are identical to what would be obtained from a
modified commutator of $\hat{q}$ and $\hat{p}$ as in (\ref{ModComm}),
$[\hat{q},\hat{\tilde{p}}]= i\hbar(1-2\pi^2\hat{\tilde{p}}{}^2/p_0^2)$
with $\tilde{p}:=p_0(\hat{h}-\hat{h}^{\dagger})/2\pi$, as it follows
from a formal operator expansion
\[
 \left[\hat{q},\hat{h}-\hat{h}^{\dagger}\right]=
 \left[\hat{q},\frac{4\pi i\hat{p}}{p_0}- \frac{8\pi^3i\hat{p}^3}{3p_0^3}
+\cdots\right]=
 -\frac{4\pi\hbar}{p_0} \left(1-
 \frac{2\pi^2}{p_0^2}\hat{p}^2+\cdots\right)\,.
\]
 (This contribution to the corrected uncertainty relation for
  systems with compact configuration space is analogous to what is
  discussed in \cite{ModUncert}.)

However, the moments on the left-hand side of the uncertainty relation
provide additional corrections to this order which must be included
for a consistent expansion. Generalized uncertainty principles thus
are not just consequences of modified commutators. We will need
$\Delta(qh)$, ${\rm Re}\Delta h^2$ and $\Delta(hh^*)$ up to the order
$p_0^{-4}$:
\begin{eqnarray}
 \Delta(hh^*) &=& \frac{4\pi^2}{p_0^2}\Delta p^2- \frac{4\pi^4}{3p_0^4}\Delta(p^4)+
 \frac{4\pi^4}{p_0^4}(\Delta p^2)^2+ \frac{8\pi^4}{p_0^4}p^2\Delta p^2\\
 \Delta h^2 &=& -\frac{4\pi^2}{p_0^2}\Delta p^2- \frac{8\pi^3i}{p_0^3}\Delta(p^3)- 
\frac{16\pi^3i}{p_0^3}p
 \Delta p^2\nonumber\\
&&+ \frac{28\pi^4}{3p_0^4}\Delta(p^4)+ \frac{32\pi^4}{p_0^4}p
 \Delta(p^3)- \frac{60\pi^4}{p_0^4}(\Delta p^2)^2- \frac{24\pi^4}{p_0^4}p^2\Delta p^2\\
 \Delta(qh) &=& \frac{2\pi i}{p_0}\Delta(qp)- \frac{2\pi^2}{p_0^2} \Delta(qp^2)-
 \frac{4\pi^2}{p_0^2}p\Delta(qp)- \frac{4\pi^3i}{3p_0^3}\Delta(qp^3)-
 \frac{4\pi^3i}{p_0^3}p\Delta(qp^2)- \frac{4\pi^3i}{p_0^3}p^2\Delta(qp)\nonumber\\
 &&+
 \frac{2\pi^4}{3p_0^4} (\Delta(qp^4)+4p\Delta(qp^3)+6p^2\Delta(qp^2) +4p^3\Delta(qp))\,.
\end{eqnarray}
A demonstration of the lengthy calculations can be found in
Appendix~\ref{a:Example}.  Moreover,
\begin{equation}
 \Delta(h^2h^*)= \frac{8\pi^3 i}{p_0^3} \Delta(p^3)- \frac{8\pi^4}{p_0^4}
 \left(\Delta(p^4)+ 2p\Delta(p^3)- 7(\Delta p^2)^2- 6p^2\Delta p^2\right)\,.
\end{equation}
(One can verify that the reality condition (\ref{RealityIV}) is
identically satisfied in terms of the $(q,p)$-moments.)

To this order, our three uncertainty relations read
\begin{eqnarray}
&& \Delta q^2\Delta(p^4)+4p\Delta q^2\Delta(p^3)- 7\Delta q^2(\Delta p^2)^2- 2p^2\Delta q^2\Delta p^2 \nonumber\\
&& - 4p\Delta(qp)\Delta(qp^2) - \Delta(qp^2)^2- 4p^2\Delta(qp)^2 
\geq \hbar^2 p^2 \,,
 \label{UncertFourthI}
\end{eqnarray}
from (\ref{UncertI}),
\begin{eqnarray}
&&\Delta q^2\Delta p^2- \Delta(qp)^2- \frac{4\pi^2}{3p_0^2}
 \left(\Delta q^2\Delta(p^4)+3 p\Delta q^2\Delta(p^3)-6\Delta q^2(\Delta p^2)^2- 3p^2\Delta q^2\Delta p^2\right)\nonumber\\
&&-\frac{4\pi^2}{3p_0^2}
 \left(
- \Delta(qp)\Delta(qp^3)- 3p\Delta(qp)\Delta(qp^2)-
 3p^2\Delta(qp)^2\right)
\geq \frac{\hbar^2}{4}
 \left(1-4\pi^2\frac{p^2+\Delta p^2}{p_0^2}\right)\label{UncertFourthII} 
\end{eqnarray}
from (\ref{UncertII}), 
and
\begin{equation}
\Delta p^2\Delta(p^4)- \Delta(p^3)^2-
 7(\Delta p^2)^3- 6p^2(\Delta p^2)^2\geq 0 \label{UncertFourthIII}\,.
\end{equation}

In order to eliminate some of the high-order moments in terms of
second-order ones, we rewrite the three uncertainty relations as
follows: (\ref{UncertFourthIII}) implies
\begin{equation}
 \Delta_1:=\Delta(p^4)-7(\Delta p^2)^2-6p^2\Delta p^2\geq 
\frac{\Delta(p^3)^2}{\Delta p^2}\geq 0
\end{equation}
while (\ref{UncertFourthI}) can be written as
\begin{eqnarray}
 \Delta_2&:=& \Delta q^2\left(\Delta(p^4)-7(\Delta p^2)^2-6p^2\Delta p^2\right)+
 4p\Delta q^2\Delta(p^3) \nonumber\\
 &&+4p^2\left(\Delta q^2\Delta p^2-\Delta(qp)^2- \hbar^2/4\right)-
 4p\Delta(qp)\Delta(qp^2)\geq \Delta(qp^2)^2\geq 0\,.
\end{eqnarray}
With the two non-negative quantities $\Delta_1$ and $\Delta_2$, the central
uncertainty relation (\ref{UncertFourthII}) reads
\begin{eqnarray}
 && \Delta q^2\Delta p^2-\Delta(qp)^2  \geq\frac{\hbar^2}{4}
 \left(1-4\pi^2\frac{p^2+\Delta p^2}{p_0^2}\right)\nonumber\\
&&\quad + \frac{\pi^2}{p_0^2}
 \left(\Delta_2+\frac{1}{3}\Delta q^2\Delta_1+
 \hbar^2p^2+\frac{4}{3}\Delta q^2(\Delta p^2)^2-
 \frac{4}{3}\Delta(qp)\Delta(qp^3)\right)\nonumber\\
&\geq& \frac{\hbar^2}{4}\left(1-\frac{4\pi^2}{p_0^2}\left(\Delta p^2
 +\frac{4}{3}\frac{\Delta q^2(\Delta p^2)^2}{\hbar^2}- \frac{4}{3}
 \frac{\Delta(qp)\Delta(qp^3)}{\hbar^2}\right)\right) \label{GUP1}
\end{eqnarray}
using $\Delta_1\geq 0$ and $\Delta_2\geq 0$ (and $\Delta q^2\geq 0$) in the
last step. If we assume that $\Delta(qp)=0$, only the remaining two
fluctuations appear; all higher moments have been eliminated to order
$p_0^{-4}$ in favor of additional fluctuation terms. Moreover, we can
self-consistently insert the uncertainty relation on its right-hand side in
(\ref{GUP1}) to bound $\Delta q^2\Delta p^2$ from below, resulting in the
generalized uncertainty relation
\begin{equation}
\Delta q^2\Delta p^2\geq \frac{\hbar^2}{4}\left(1-\frac{16\pi^2}{3p_0^2} 
\Delta p^2\right)
\end{equation}
expanded to second order in $1/p_0$. Taking a square root to this
order, we have
\begin{equation}\label{GUPDerived}
 \Delta q\Delta p\geq  \frac{\hbar}{2} \left(1- \frac{8\pi^2}{3}
 \frac{(\Delta p)^2}{p_0^2}\right)
\end{equation}
which is of the form (\ref{GUP}) with a negative
$\alpha=-8\pi^2/3p_0^2$. We see that $\Delta q$ can vanish at a finite
critical value of $\Delta p_c$, namely $\Delta p_c = \sqrt{-1/\alpha}=
\sqrt{3/2}p_0/2\pi$. While this value for $\Delta p_c$ shows the
expected qualitative behavior, it can only be a rough estimate, given
that the correction term $8\pi^3 (\Delta p_c)^2/3p_0$ is certainly not
small when it cancels the standard term $\hbar/2$ of the uncertainty
relation.  Nevertheless, the so-obtained value for the critical
$\Delta p_c$ is quite close to what we derived earlier for a position
eigenstate.  Our expansion by the moments assumes that all momentum
variables, including the moments, are small compared to suitable
powers of $p_0$, with $\Delta(p^n)/p_0^n$ falling off as $n$ gets
larger.  Even for $n=2$, the ratio is not small compared to one. For
higher moments, as remarked at the end of Sec.~\ref{s:Periodic},
position eigenstates (corresponding to $\Delta q=0$) do fulfill the
fall-off assumption, but with a comparatively small rate of
$2^{-n}/(n+1)$. (For comparison, semiclassical expansions usually make use
of moments falling off as $\hbar^{n}$ relative to some classical scale
with the dimension of an action, providing much smaller numbers.)
Leaving position eigenstates aside, there is a large class of states
that easily fulfill our assumptions provided they are sufficiently
strongly peaked in $p$.  For such states, our generalized uncertainty
relation (\ref{GUPDerived}) reliably exhibits implications of discrete
space on fluctuations.

\section{Conclusions}

We have derived the first order of corrections to the standard
uncertainty relation as they result for a quantum system with a
momentum space of the topology of $S^1$ and thus discrete position.
Without needing to assume corrections to the basic operator algebra
(\ref{algebra}), we showed that an underlying discreteness of position
spectra implies specific respresentation-independent correction terms
in a generalized uncertainty principle. Formally, there is no
self-adjoint operator associated with the coordinate of the compact
direction of the phase space, which is rather quantized via periodic
functions of an angular coordinate.  (Group-theoretical quantization
\cite{Isham}, for instance, can be used to construct the quantum
representation.) For angle separations small compared to the
periodicity one can then expand quantum variables such as
fluctuations, correlations and higher moments and, to leading order,
reproduce the standard uncertainty relations. Higher orders of the
expansion, which include terms sensitive to the periodicity, lead to a
derived form of a generalized uncertainty principle.

Heuristically, a generalized uncertainty principle of a form that implies a
positive lower bound for position uncertainty has been interpreted as a signal
of spatial discreteness, as it may be realized in quantum gravity. This has
been supported in \cite{NonLocalGUP} by an analysis of the representation
theory of operator algebras which imply such a generalized uncertainty
principle. Perhaps surprisingly, the specific form of the generalized
uncertainty principle derived in our calculations has the opposite sign of its
coefficients compared to what leads to a finite minimal position uncertainty:
Even though we know that the underlying Hilbert space implies discrete spectra
and thus spatial discreteness in a rigorous sense, there is no finite lower
bound to $\Delta q$.

Of course, as we discussed, one may expect the absolute minimum to be zero
because normalizable eigenstates of sharp position exist. In this case, a more
refined version of minimum uncertainty can be introduced which depends on the
expectation value $\langle\hat{q}\rangle$: the minimum uncertainty could
vanish when $\langle\hat{q}\rangle$ equals an eigenvalue of $\hat{q}$, but
would be non-zero otherwise. Such relations for the minimum $\Delta q_{\rm
  min}(\langle\hat{q}\rangle)$ can be derived at the Hilbert space level, but
are not realized by the treatment used here. As we showed in Section
\ref{s:Periodic}, the presence of non-vanishing minima of fluctuations depends
on the quantum representation. Generalized uncertainty principles, on the
other hand, are representation independent as derived here; they follow from
algebraic properties of quantum observables. While leading corrections to the
standard uncertainty relation are $\langle\hat{q}\rangle$-independent and
  cannot directly give rise to minimal uncertainties of the functional form
  $\Delta q_{\rm min}(\langle\hat{q}\rangle)$, one may
expect that higher orders could bring in such a dependence on
$\langle\hat{q}\rangle$. Indeed, the dependence of $\Delta q_{\rm min}$ on
$\langle\hat{q}\rangle$ is most pronounced near $\hat{q}$-eigenstates, where
the leading terms of the expansion in moments are not reliable.  If higher
orders are included, such a dependence may arise at least indirectly via
moments involving $q$. These moments are independent of the expectation value,
but specific classes of states, such as $\hat{q}$-eigenstates, could imply
restrictions on the moments compatible with the form of $\Delta q_{\rm min}$
seen before in (\ref{Deltaqmin}). We leave this question open for future
investigations.

Thus, there is no simple relationship between positive lower bounds
for uncertainties according to generalized uncertainty principles on
one hand, and true discreteness of operator spectra on the underlying
Hilbert space on the other. One may view the existence of a positive
lower bound for $\Delta q$ as an indication for a theory with a
universal bandwidth, or a theory based on extended fundamental
objects, which would be consistent with the fact that generalized
uncertainty relations with a positive lower bound have been argued to
arise, also from string theory. A key signature of a fundamental
discreteness of space, by contrast, is the possibility of vanishing
position fluctuations at finite momentum fluctuation. We re-emphasize,
however, that our treatment works well for values of variables small
compared to their periodicity, for which curvature bounds in quantum
gravity are an example. If one instead probes an underlying periodic
structure of position space, separations comparable to the periodicity
scale would have to be considered where our present expansions do not
apply.

As an alternative to string theory as a quantum theory of gravity,
loop quantum gravity \cite{Rov,ALRev,ThomasRev} provides a kinematical
quantization where geometrical operators have discrete spectra
\cite{AreaVol,Vol2}. While this property has not been derived for
physical observables, the discrete form of kinematical spectra affects
the dynamics because of the form of basic operators which are combined
to a Hamiltonian (constraint) operator. Dynamical implications can be
studied in loop quantum cosmology \cite{LivRev,Springer,SIGMA}, for
instance in the context of space-time singularities \cite{BSCG}. The
formulation of isotropic models in loop quantum cosmology makes use of
complex exponentials of curvatures, rather than curvature components
themselves \cite{IsoCosmo}. The example analyzed here can thus be
taken as a model for isotropic loop quantum cosmology, which indicates
the form of generalized uncertainty principles as they may appear in
cosmological applications. Our results here would apply only to
small-curvature regimes where the discreteness of spatial geometry
does not play a large role, corresponding to the fact that we had to
expand our exponentials on a circle in the inverse periodicity in
order to derive our generalized uncertainty principle.

Taking the circle example as a model for the kinematical structure of a sector
in loop quantum cosmology suggests that the canonical variables $V$ and $P$,
related to the volume and expansion rate as introduced in
Appendix~\ref{a:LQC}, are subject to a generalized uncertainty principle
\begin{equation}
 \Delta V\Delta P\geq \frac{\hbar}{2} \left(1- 
\frac{2}{3}(\Delta P)^2\right)\,.
\end{equation}
This inequality is valid as long as $P$ and $\Delta P$ are small
compared to the scale $P_0=2\pi$ of almost periodicity.  (As in the
general derivation, we also assume a vanishing $(V,P)$-covariance;
otherwise there will be additional corrections as shown by the
previous formulas.)  Loop quantum cosmology does not show uniquely
what variables behave almost-periodically. Taking ambiguities into
account, the periodicity scale in terms of the scale factor is set by
two parameters $f_0$ and $x$ according to the power-law
parameterization $P=-f_0a^{2x}\dot{a}$. The dimension of $f_0$ depends
on the value of $x$, given that $P$ must be dimensionless. For the
value $x=-1/2$, for instance, $f_0$ has the dimension of length and
due to its quantum-gravity origin one may expect it to be of the order
of the Planck length $f_0\sim \ell_{\rm P}=\sqrt{G\hbar}$. (For
consistency with other corrections from loop quantum cosmology, it
must be sufficiently larger than the Planck length \cite{Consistent}.)
In this case, a Planckian bound $\dot{a}/a<\ell_{\rm P}^{-1}$ for the
Hubble parameter is required for the applicability of our derivations
here and leading corrections are of the order $(\ell_{\rm
P}\Delta(\dot{a}/a))^2$.

In fact, as observed in \cite{LoopGUP}, the use of modified commutation
relations between the canonical variables which correspond to a
generalized uncertainty principle of the form derived here can mimic
some of the effects of loop quantum cosmology. The main example is a
bounce in isotropic models sourced by a free scalar
\cite{QuantumBigBang,BouncePert}. However, such an example for
high-curvature effects appears when $P\sim P_0$ and thus falls outside
the regime where derivations of the present paper are valid. We
nevertheless note that our derivations are not restricted to purely
semiclassical regimes; all we need is a hierarchy of moments organized
by powers of $P_0^{-1}$, not of $\hbar$.

In addition to the gravitational degrees of freedom, loop quantization
also applies to matter fields. A scalar field, for instance, can be
represented on the loop Hilbert space in an almost-periodic fashion
similar to the gravitational connection or the canonical variable $P$
in isotropic cosmology \cite{ScalarBohr,ExpScalar}. In a setting of
quantum field theory, generalized uncertainty relations should then
appear, with possible phenomenological consequences during inflation.

We conclude by emphasizing again that our considerations here were
kinematical, using a moment expansion in uncertainties. The same tool
is the key to analyzing quantum back-reaction effects in the dynamics,
where equations of motion (or constraints) are expanded by moments
\cite{EffAc}. This can be done either in canonical
variables or in variables analogous to $h$ used on the circle
\cite{BouncePert}. We leave it open to
further studies to see what a combination of both types of moment
expansions would provide.

\section*{Acknowledgements}

MB acknowledges partial support by NSF grant PHY0748336.  AK
acknowledges support from the Canada Research Chairs and Discovery
programs of the National Science and Engineering Research Council of
Canada (NSERC).

\begin{appendix}

\section{Loop quantum cosmology}
\label{a:LQC}

 We present a brief review of loop quantum cosmology with a focus on
  aspects relevant for questions of the discreteness or periodicity of some
  directions in phase space. In this context, we must take a general viewpoint
  in order to see all possible forms of discreteness that can arise,
  especially at a dynamical level. Our summary here therefore differs from
  some contributions and reviews in the recent literature, where models are
  specialized further by ad-hoc choices so as to produce detailed studies of
  some specific cases.

In loop quantum gravity \cite{Rov,ALRev,ThomasRev}, one uses as one of
the basic canonical fields a densitized triad $E^a_i$ of three
orthonormal vector fields labelled by $i=1,2,3$, related to the
spatial metric $q_{ab}$ by $E^a_iE^b_i=\sqrt{\det q}q^{ab}$. As a
smeared version, the field is quantized via flux operators
$\hat{F}(S)=\int_S\hat{E}^a_in_a{\rm d}^2y$ integrated over
2-dimensional surfaces in space rather than by its pointwise
values. In an isotropic setting, $E^a_i=p\delta^a_i$ is completely
determined by the scale factor $a$ up to orientation, with $|p|=a^2$
and the sign of $p$ giving the orientation of space. Fluxes, then,
reduce to area-like quantities such as $A=\ell_0^2|p|$ where $\ell_0$
provides a linear measure (in terms of coordinates) for the surfaces
used. 

In quantum states, areas $A$ obtained from flux operators play the role of
quantum numbers that determine the elementary discreteness of space.  Indeed,
the quantum representation implies a discrete spectrum for flux operators,
whose smallest possible non-zero values are of the order $A\sim \ell_{\rm
  P}^2$. One is thus led to a discrete (mini\-super)space as used in this
article. For isotropic geometries, the canonically conjugate almost-periodic
momentum of $A$ is $\ell_0\dot{a}$ (represented via holonomy operators). But
while the spectrum of flux operators {\em for fixed surfaces} is fully
determined and of a simple equidistant form, the question of what the
dynamical stepsize of {\em physical scales} is, for instance in an expanding
universe, remains open. The dynamics of a classical expanding universe is
described by the scale factor or the triad variable $p$, while elementary
fluxes in quantum theory determine the possible sizes of $\ell_0^2|p|$ with
$\ell_0$ depending on the coordinate size of surfaces (or plaquettes in a
lattice-like state of discrete space) giving rise to the smallest flux
eigenvalues. If the lattice is changing, a process called lattice refinement
which is generically realized in loop quantum gravity
\cite{InhomLattice,CosConst}, $\ell_0$ must be assumed to depend on time or
the scale factor as well. The known equidistant spectrum for fluxes $A$ then
determines the stepsize of geometrical measures related to the scale factor
only if $\ell_0$ for lattice plaquettes is known as a function of $a$ or $p$.

Evaluating the full dynamics of loop quantum gravity, for instance as
in \cite{QSDI}, remains extremely challenging; it is thus impossible
to derive some function $\ell_0(p)$ from first principles. However, on
general grounds there are certain restrictions on its behavior. If
$\ell_0$ did not depend on $p$, for instance, the discreteness scale
of a lattice state would be constant in terms of coordinates, but
would be magnified as the scale $a\ell_0$ measured in an expanding
universe. For sufficiently long expansion, one would be in conflict
with continuum physics. A decreasing scale $\ell_0(p)$ is thus
required, one useful example being the power-law form
$\ell_0(p)=f_0|p|^x$ with two constants $f_0$ for the discreteness
scale and $x<0$ for the refinement behavior.  It is then the product
$\ell_0(p)\dot{a}=f_0a^{2x}\dot{a}$, not $\dot{a}$, which is almost
periodic, and the conjugate variable $\int\ell_0(p)^{-1}{\rm d}p=
f_0^{-1}|p|^{1-x}/(1-x)$, not $p$, which is equidistant.

In terms of the cosmological scale factor $a$, we thus define canonical
variables
\begin{equation} \label{LoopVars}
 V=\frac{3\sigma {\cal V}a^{2-2x}}{8\pi  G(1-x)f_0} \quad\mbox{ and }\quad 
P=-f_0a^{2x}\dot{a} \quad\mbox{with}\quad \{V,P\}=1
\end{equation}
where $G$ is the gravitational constant. These conventional variables absorb
the precise periodicity scale of $a^{2x}\dot{a}$ in $f_0$ such that $P_0=2\pi$
and $\mu_n^{(\epsilon)}=n+\epsilon/2\pi$.  In $V$, moreover, the spatial
volume ${\cal V}$ of an integration region used to average to isotropy,
measured in coordinates, appears, as well as $\sigma=\pm1$ which determines
the orientation of space. With the factor of $\sigma$, allowed values of $V$
cover the whole real line because loop variables are derived from triads,
which by changing orientation can take both signs; see \cite{Springer} for
derivations and details.

The dynamics of a loop quantum cosmological model takes different forms
depending on which variable precisely is almost periodic. Unlike the
condensed-matter example in Section \ref{s:Basis}, it is not clear a priori
whether it is, say, $a$ itself which acquires an equidistant spectrum in any
of the periodic dynamical sectors, or a different power of $a$ (or yet another
functional behavior). We therefore keep this freedom in our definition of
basic variables where the power $x$ remains unspecified. (Arguments loosely
based on the full theory of loop quantum gravity indicate that $-1/2<x<0$
generically \cite{InhomLattice,CosConst}, with values near $-1/2$ preferred
phenomenologically \cite{RefinementInflation,RefinementMatter,ScalarHol} at
least in near-isotropic cosmology.)  Moreover, even if the precise discrete
variable would be specified, the discreteness scale remains free. This is
parameterized by the second constant $f_0$ whose dimension depends on
$x$.\footnote{In this context, one should be careful due to the scaling
  behavior of $a$: as the scale factor of a universe, its value changes
  whenever spatial coordinates are rescaled by a constant. Thus, $f_0$ is not
  coordinate independent (unless $x=-1/2$) because it must absorb the
  coordinate dependence of the scale factor. For a similar reason, $f_0$
  depends on the spatial averaging volume ${\cal V}$ which is not only
  coordinate dependent but also changes whenever a different integration
  region is chosen. Both dependences of $f_0$ are only artefacts of the
  isotropic formulation in terms of the scale factor, and do not imply that
  the physical model would depend on coordinates or the choice of an
  integration region. (The scaling issue is not altogether avoided in
  spatially closed models with their compact total space because one may still
  use regions smaller than the total space for averaging to isotropy.)}

A further difference to the Bloch example is that this so-called kinematical
Hilbert space of states (\ref{Pstates}), as it follows\footnote{There are
  different ways to derive the basic representation on the kinematical Hilbert
  space from full loop quantum gravity
  \cite{SymmRed,InhomLattice,SymmStatesInt,Reduction,Rieffel}. The derivation
  of dynamics via a reduced Hamiltonian constraint operator from a full one
  is, however, more complicated. If this could be done in sufficient detail,
  the parameters $f_0$ and $x$ could in principle be determined.}  from the
full theory of loop quantum gravity, carries a different representation than
is typically used in quantum mechanics \cite{Bohr}: All states
$\psi_n^{(\epsilon)}$ are normalizable despite their plane-wave form, and they
form an orthonormal basis. (Although non-standard, this representation may be
advantageously used also in quantum mechanics \cite{BohrQM} and quantum field
theory \cite{QEDBohr}.) Since there are uncountably many such states, the
Hilbert space is non-separable. A specific way to write the inner product is
the integral form
\begin{equation}
 \langle f,g\rangle= \lim_{T\to\infty} \frac{1}{2T}\int_{-T}^T 
\overline{f(P)}g(P){\rm d} P\,.
\end{equation}

Since $V$ is conjugate to $P$, it can be represented as the usual
derivative operator $\hat{V}=i\hbar \partial/\partial P$. The states
(\ref{Pstates}) then turn out to be true normalizable eigenstates of
$\hat{V}$, which thus has a discrete spectrum. For the scale factor
$a$, the eigenvalues in terms of the quantum number $\mu_n^{(\epsilon)}$
read
\begin{equation}
 a_n^{(\epsilon)}= \left(\frac{8\pi G\hbar
 f_0(1-x)|\mu_n^{(\epsilon)}|}{3{\cal V}} \right)^{1/(2-2x)}= 
\left(\frac{8\pi G\hbar
 f_0(1-x)|n+\epsilon/2\pi|}{3{\cal V}} \right)^{1/(2-2x)}\,.
\end{equation}

As in the case of Bloch states, it is the dynamics which must
determine the specific realization and effects of the underlying
discreteness as well as potentially observable
implications. Classically, cosmological dynamics is governed by the
Friedmann equation
\begin{equation} \label{ClassConstr}
 0=C=a\dot{a}^2- \frac{8\pi G}{3}E(a)
\end{equation}
where $a$ is the scale factor and $E$ the matter energy in the
universe. Since $\dot{a}$, according to (\ref{LoopVars}) is related to
the variable $P$ which, after a loop quantization, becomes almost
periodic, it is not possible to represent the Friedmann equation
directly on the Hilbert space of loop quantum cosmology. Instead, one
has to look for an operator which is well-defined and which produces
$\dot{a}^2$ in the classical limit of small curvature where
$\dot{a}\ll 1$ (or more precisely $f_0a^{2x}\dot{a}\ll 1$). With
$P$ parameterized to reflect the scale of almost periodicity, a simple
and often-used operator that satisfies the requirements is obtained
after replacing $a\dot{a}^2$ in (\ref{ClassConstr}) with
$f_0^{-2}a^{1-4x}\sin^2P$ where $a^{1-4x}$ is proportional to
$V^{2-3/(2-2x)}$ in terms of canonical variables. This specific
process of adapting the classical equation in large-curvature regimes
is called ``holonomy modification.'' It plays the role of a
regularization to ensure that the classical expression can be promoted
to an operator in the quantum representation used.

A detailed derivation of the precise functional form of the Hamiltonian, or
the specific form of functions such as $\sin^2P$ in holonomy modifications,
must await further developments in evaluating the theory. This would be like
asking to derive the potential $V(x)$ relevant for the motion of electrons in
a crystal from first principles of the underlying many-body system composed of
all nuclei and electrons. Such a derivation is certainly complicated, but
still the Hamiltonian resulting from the simple basic assumptions made above
has several characteristic properties for which the detailed form is not
crucial. They influence the dynamics, which in qualitative terms will depend
on the size of parameters such as $f_0$ and $x$. In contrast to a
condensed-matter Hamiltonian, in this context one is not interested in all
energy eigenvalues but only in the zero eigenspace, so-called physical states
annihilated by the combined Hamiltonian of gravity and matter which forms a
constraint rather than an expression of energy. There is thus no band
structure, but implications of the discreteness do show up in other dynamical
properties of the solutions.

From the action of a holonomy modification like $\sin^2P$ as a
multiplication operator
\begin{equation}
 \widehat{\sin^2P} \psi_n^{(\epsilon)}(P)= -\frac{1}{4}\left(
\psi_{n+2}^{(\epsilon)}(P)- 2\psi_n^{(\epsilon)}(P)+ \psi_{n-2}^{(\epsilon)}(P)
\right)
\end{equation}
on $\hat{V}$-eigenstates $\psi_n^{(\epsilon)}$ of the form (\ref{Pstates}),
with a matter Hamiltonian operator $\hat{E}\psi_n^{(\epsilon)}(P)=
E_n^{(\epsilon)} \psi_n^{(\epsilon)}(P)$, the constraint $C=0$ in
(\ref{ClassConstr}) is quantized to a difference equation
\cite{cosmoIV,IsoCosmo}
\begin{equation} \label{QuantConstr}
 C_+^{(\epsilon)}(n)s_{n+2}^{(\epsilon)}+ C_0^{(\epsilon)}(n)s_n^{(\epsilon)}+ 
C_-^{(\epsilon)}(n)s_{n-2}^{(\epsilon)}=\frac{8\pi G}{3} E_n^{(\epsilon)} 
s_n^{(\epsilon)}
\end{equation}
for the coefficients of physical states $\psi(P)= \sum_{n,\epsilon}
s_n^{(\epsilon)}\psi_n^{(\epsilon)}(P)$ expanded in (\ref{Pstates}). The
coefficients $C_0^{(\epsilon)}(n)$ and $C_{\pm}^{(\epsilon)}(n)$ of the
difference equation follow from quantizing the $a$-dependent terms in
(\ref{ClassConstr}); see e.g.\ \cite{IsoCosmo,Sing,Springer} for concrete
examples. Eq.~(\ref{QuantConstr}) may appear like an eigenvalue equation for
$E_n^{(\epsilon)}$, but solutions to this constraint are not required to be
normalizable. In fact, if the system describes an ever-expanding cosmology,
wave functions are expected to be supported at all $n$ without a strong
fall-off at $n\to\pm\infty$.  Thus, general solutions are not
normalizable. However, they describe the change of the wave function of an
evolving universe for any given $\hat{E}$ in accordance with the matter model.

\section{Example for the expansion of moments}
\label{a:Example}

Here we show some of the calculations necessary to expand moments up
to third order in $p_0^{-1}$. In the main text, we had to use results
up to fourth order, which are more lengthy but follow from analogous
calculations.

First, we have
\begin{equation}
 \Delta h^2= \langle\hat{h}^2\rangle-h^2
 = -\frac{4\pi^2}{p_0^2}\Delta p^2 - \frac{8\pi^3i}{p_0^3}\Delta(p^3)- 
\frac{16\pi^3i}{p_0^3}p \Delta p^2
+\cdots\,.
\end{equation}
where we used the third-order moment
\begin{equation}
 \Delta(p^3)= \langle(\hat{p}-p)^3\rangle= \langle\hat{p}^3\rangle-
 3p\langle\hat{p}^2\rangle+ 2p^3\,.
\end{equation}

For mixed moments we have to be more careful with the ordering:
\begin{eqnarray}
 \Delta(qh)&=&\frac{1}{2}
 \langle\hat{q}\hat{h}+\hat{h}\hat{q}\rangle-qh\\
 &=& \frac{i\pi}{p_0} \langle\hat{q}\hat{p}+\hat{p}\hat{q}\rangle
 -\frac{2\pi i}{p_0}qp-
 \frac{\pi^2}{p_0^2}\langle\hat{q}\hat{p}^2+\hat{p}^2\hat{q}\rangle+
 \frac{2\pi^2}{p_0^2} q\langle\hat{p}^2\rangle-
 \frac{2\pi^3i}{3p_0^3}\langle\hat{q}\hat{p}^3+\hat{p}^3\hat{q}\rangle+
 \frac{4\pi^3i}{3p_0^3}q\langle\hat{p}^3\rangle+\cdots \nonumber\\
 &=& \frac{2\pi i}{p_0}\Delta(qp)- \frac{2\pi^2}{p_0^2} \Delta(qp^2)+
 \frac{4\pi^2}{p_0^2}p\Delta(qp)- \frac{4\pi^3i}{3p_0^3}\Delta(qp^3)-
 \frac{4\pi^3i}{p_0^3}p\Delta(qp^2)-
 \frac{4\pi^3i}{p_0^3}p^2\Delta(qp)+\cdots\nonumber 
\end{eqnarray}
where in the last step the moments
\begin{eqnarray}
 \Delta(qp^2) &=& \frac{1}{3}\langle(\hat{q}-q)(\hat{p}-p)^2+
 (\hat{p}-p)(\hat{q}-q)(\hat{p}-p)+ (\hat{p}-p)^2(\hat{q}-q)\rangle\nonumber\\
 &=& \frac{1}{2}\langle\hat{q}\hat{p}^2+\hat{p}^2\hat{q}\rangle-
 q\Delta p^2-2p\Delta(qp)-qp^2\\
 \Delta(qp^3) &=& \frac{1}{4} \langle(\hat{q}-q)(\hat{p}-p)^3+
 (\hat{p}-p)(\hat{q}-q)(\hat{p}-p)^2+ 
 (\hat{p}-p)^2(\hat{q}-q)(\hat{p}-p)+
 (\hat{p}-p)^3(\hat{q}-q)\rangle\nonumber\\
 &=& \frac{1}{2}\langle\hat{q}\hat{p}^3+\hat{p}^3\hat{q}\rangle-
 q\Delta(p^3)- 3p\Delta(qp^2)- 3p^2\Delta(qp)-qp^3
\end{eqnarray}
have been used.

\end{appendix}


\begin{thebibliography}{10}
\bibitem{ShortDistanceThree}
A.\ Kempf,
\newblock On the only three Short Distance Structures which can be described by
  Linear Operators,
\newblock {\em Rept.\ Math.\ Phys.} 43 (1999) 171--177, [hep-th/9806013]
\end{thebibliography}

\end{document}